\documentclass[twocolumn, 10pt]{IEEEtran} %!PN
   \usepackage{color}
   \usepackage{latexsym}
   \usepackage{pslatex}
   \usepackage{algorithm}
   \usepackage{algorithmic}
   \usepackage{amsmath}
   \usepackage{balance}

\begin{document}

\title{AIG Rewriting Using 5-Input Cuts} 
\author{Nan~Li \hspace{24mm} Elena~Dubrova\\
Royal Institute of Technology, ES/ICT/KTH, 164 46 Kista, Sweden}
\maketitle

% remove page numbers
%\thispagestyle{empty}
%\pagestyle{empty}

%------------------------------------------------------------------------- 
\begin{abstract}
Rewriting is a common approach to logic optimization based on local transformations.
Most commercially available logic synthesis tools include a rewriting engine that may be used
multiple times on the same netlist during optimization.
This paper presents an And-Inverter graph based rewriting algorithm using 5-input cuts. 
The best circuits are pre-computed for a subset of NPN classes of 5-variable functions.
Cut enumeration and Boolean matching are used to identify replacement candidates.
The presented approach is expected to complement existing rewriting approaches 
which are usually based on 4-input cuts. 
The experimental results show that, by adding the new rewriting algorithm to ABC synthesis tool, 
we can further reduce the area of heavily optimized large circuits by 5.57\% on average.
\end{abstract}
%------------------------------------------------------------------------- 


%------------------------------------------------------------------------- 
\section{Introduction}	\label{sec:intro}

Logic optimization approaches can be divided into {\em algorithmic-based methods}, which are based on global transformations, and {\em rule-based methods}, which are based on local transformations~\cite{espr}.
Rule-based methods, also called {\em rewriting},  use a set of rules which are applied when certain patterns are found. A rule transforms a pattern for a local sub-expression, or a sub-circuit, into another equivalent one. Since rules need to be described, and hence the type available of operations/gates must be known, the rule-based approach usually requires that the description of the logic is confined to a limited number of operation/gate types such as AND, OR, XOR, NOT etc. In addition, the transformations have limited optimization capability since they are local in nature. Examples of rule-based systems include LSS~\cite{lss} and SOCRATES~\cite{socrates}.

Algorithmic methods use global transformations such as decomposition or factorization, and therefore they are much more powerful compared to the rule-based methods. However, general Boolean methods, including don't care optimization, do not scale well for large functions. Algebraic methods are fast and robust, but they are not complete and thus often give lower quality results. For this reasons, industrial logic synthesis systems normally use algebraic restructuring methods in a combination with rule-based methods.

In this paper, 
we propose a new rewriting algorithm based on 5-Input cuts. In the algorithm, the best circuits are pre-computed for a subset of NPN classes of 5-variable functions. Cut enumeration technique~\cite{Cong99} is used to find 5-input cuts for all nodes, and some of them are replaced with a best circuit. The Boolean matcher~\cite{Chai06} is used to map a 5-input function to its canonical form.
The presented approach is expected to complement existing rewriting approaches 
which are usually based on 4-input cuts. 
Our experimental results show that, by adding the new rewriting algorithm to ABC synthesis tool~\cite{abc}, 
we can further reduce the area of heavily optimized large circuits by 5.57\% on average.

The paper is organized as follows.
Section~\ref{sec:bg} describes main notions and definitions used in the sequel.  
Section~\ref{sec:prev} summarises previous work.
Section~\ref{sec:main} presents the proposed approach.
Section~\ref{sec:exp} shows experimental results. Section~\ref{sec:conc} concludes the paper and
discusses open problems.

%------------------------------------------------------------------------- 
\section{Background} \label{sec:bg}

A \emph{Boolean network} is a directed acyclic graph, of which the nodes represent logic gates, and the directed edges represent connections of the gates. A network is also referred to as a \emph{circuit}.

A node of the network has zero or more \emph{fanins}, and zero or more \emph{fanouts}. A \emph{fanin} of a node $n$ is a node $n_\text{in}$ such that there exists an edge from $n_\text{in}$ to $n$. Similarly, a \emph{fanout} of a node $n$ is a node $n_\text{out}$ such that there is an edge from $n$ to $n_\text{out}$. The \emph{primary inputs} (PIs) of a network are the zero-fanin nodes of the network. The \emph{primary outputs} of a network are a subset of all nodes. If a network contains flip-flops, the inputs/outputs of the flip-flops are treated as POs/PIs of the network.

An \emph{And-Inverter graph} (AIG) is a network, of which a node is either a PI or a 2-input AND gate, and an edge is negatable. An AIG is structurally hashed~\cite{Ganai00} to ensure uniqueness of the nodes. The \emph{area} of an AIG is measured by the number of nodes in the network.

A \emph{cut} of a node $n$ is a set $C$ of nodes such that any path from a PI to $n$ must pass through at least one node in $C$. Node $n$ itself forms a \emph{trivial cut}. The nodes in $C$ are called the $leaves$ of cut $C$. A cut $C$ is \emph{$K$-feasible} if $|C| \leq K$; additionally, $C$ is called a \emph{$K$-input cut} if $|C| = K$.

The \emph{level} of a node $n$ is the number of edges of the longest path from any PI to $n$. The \emph{depth} of a network is the largest level among all internal nodes of the network.

Two Boolean functions, $F$ and $G$, are \emph{NPN-equivalent} and belong to the same \emph{NPN equivalence class}, if $F$ can be transformed into $G$ through negation of inputs (N), permutation of inputs (P), and negation of the output (N)~\cite{hurst2}.

%------------------------------------------------------------------------- 
\section{Previous Work} \label{sec:prev}

Rewriting of networks was introduced in the early logic synthesis systems. SOCRATES~\cite{socrates} and the IBM system~\cite{lss}\cite{ibm} performed rewriting under a set of rewriting rules to replace a combination of library gates with another combination of gates which had a smaller area or delay. In SOCRATES, these rules were managed in an expert system deciding which ones to apply and when. The rules in SOCRATES were written by human designers, based on personal experience and observation of experimental results. 

In the MIS system~\cite{mis}, which later developed into SIS~\cite{sis}, local transformations such as \emph{simplification} were used to locally optimize a multi-level network after global optimization. Two-level minimization methods such as ESPRESSO~\cite{espr} were used to minimize the functions associated with the nodes in the network. Similar methods~\cite{Brayton90} were also included in works of~\cite{bold}\cite{Malik88}\cite{Savoj89}.

Rule-based rewriting method was used to simplify AND-OR-XOR networks in the multi-level synthesis approach presented in~\cite{Sasao95}.

AIG-based rewriting technique presented in~\cite{Bjesse04} is used as a way to compress circuits before formal verification. Rewriting is performed in two steps. In the first step, which happens only once when the program starts, all two-level AIG subgraphs are pre-computed and stored in a table by their Boolean functions. In the second step, the AIG is traversed in topological order. The two-level AIG subgraphs of each node are found and the functionally equivalent pre-computed subgraphs are tried as the implementation of the node, while logic sharing with existing nodes is considered. The subgraph leading to least number of overall nodes is used as the replacement of the original subgraph.

An improved AIG rewriting technique for pre-mapping optimization is presented in~\cite{rwr}. It uses 4-input cuts instead of two-level subgraphs in rewriting, and preserves the number of logic levels so the area is reduced without increasing delay. Additionally, AIG balancing, which minimizes delay without increasing area, is used together with rewriting, to achieve better results. Iterating these two processes forms a new technology-independent optimization flow, which is implemented in the sequential logic synthesis and verification system, ABC~\cite{abc}. Experiments show that this implementation scales to very large designs and is much faster than SIS~\cite{sis} and MVSIS~\cite{mvsis}, while resulting in circuits with the same or better quality.

%------------------------------------------------------------------------- 
\section{AIG Rewriting Using 5-Input Cuts} \label{sec:main}

The presented algorithm can be divided into two parts: 
\begin{enumerate}
	\item Best circuit generation \label{part:cgen}
	\item Cut enumeration and replacement \label{part:enum}
\end{enumerate}
Part \ref{part:cgen} of the algorithm tries to find the optimal circuits for a subset of ``practical'' 5-variable NPN classes, and stores these circuits. Part \ref{part:enum} of the algorithm enumerates all 5-input cuts in the target circuit, and chooses to replaces a cut with a suitable best circuit. 

In the implementation of rewriting using 4-input cuts in~\cite{rwr}, pre-computed tables of canonical forms and the transformations are kept for all $2^{16}$ 4-input functions~\cite{abc}\cite{rwr}. As we extend rewriting to 5-input cuts, the size of these tables becomes $2^{32}$. i.e. 
too large for using in a program that runs on a regular computer. In our implementation, we use a Boolean matcher~\cite{Chai06} to dynamically calculate the canonical form of a truth table and the corresponding transformation from the original truth table.

\subsection{Best circuit generation}
Similarly to~\cite{rwr}, we pre-compute the candidate circuits for each NPN class so they can be directly used later. There are $616126$ NPN equivalence classes for 5-input functions, among which only $2749$ classes appear in all IWLS 2005 benchmarks~\cite{iwls2005} as 5-feasible cuts. We picked $1185$ of them with more than 20 occurrences, and generated best circuits for representative functions of these classes. 

Due to the expanded complexity of the problem, we had to make some trade-offs between the quality of the circuits and the time and memory usage of our algorithm. Our implementation has following differences compared to~\cite{rwr}:
\begin{itemize}
	\item Use of Boolean matcher to calculate canonical form, instead of table look-up.
	\item Use of a hash map to store the candidate into best circuits, instead of using a full table.
	\item When deciding whether to store a node in the node list, a node with the same cost as an existing node is discarded, instead of being stored in the list.
	\item Nodes of both canonical functions and the complement of the canonical functions are used as the candidate circuit, while in~\cite{rwr} complement functions are not used.
	\item When the number of nodes reaches an upper limit, a reduction procedure is performed before the generation continues, leaving only the nodes used in the circuit table. 
\end{itemize}

We use two structures to store the best circuits: the \emph{forest}, list of all nodes, and the \emph{table}, storing only the pointers to the nodes in the list, which represent canonical functions or their complements. In the \emph{forest}, a node can either be an AND node or an XOR node, and two incoming edges of a node have complementation attributes. The \emph{cost} of a node is the number of AND nodes plus twice the number of XOR nodes those are reachable from this node towards the inputs.

First, the constant zero node and five nodes for single variables are added into the \emph{forest}. The constant node and one of the variable nodes are added to the \emph{table}, since all variable nodes are NPN equivalent. Then, for each pair of nodes in the \emph{forest}, five types of 2-input gates are created, using the pair as inputs:
\begin{itemize}
	\item AND gate
	\item AND gate with first input complemented
	\item AND gate with second input complemented
	\item AND gate with both inputs complemented
	\item XOR gate
\end{itemize}

A newly created node is stored in the \emph{forest} if the following conditions are met, otherwise it is discarded:
\begin{itemize}
	\item The cost of the node is lower than any other node with the same functionality.
	\item The cost of the node is lower than or equal to any other node with NPN-equivalent functionality.
\end{itemize}
In addition, the pointer to this node is added to the \emph{table} if the following condition is also met:
\begin{itemize}
	\item The function of the node is the canonical form representative, or its complement, in the NPN-equivalence class it belongs to.
\end{itemize}

When the number of nodes in the \emph{forest} reaches an upper limit, a node reduction procedure is performed, where only the reachable nodes from the nodes in the \emph{table} are left in the \emph{forest}.

The algorithm stops when the number of uncovered ``practical'' classes is smaller than a threshold value.

Finally, the generated best circuits are stored, so they can be used later when rewriting takes place.

The pseudo-code of the proposed best circuit generation algorithm is shown in Algorithm~\ref{alg:gen}. The \texttt{GenerateBestCircuits} procedure returns a node list $N$ and a table of nodes $C$ recording the candidate best circuits for a subset of NPN classes. It takes three parameters. Parameter $P$ is a set of truth tables of ``practical'' 5-variable functions. This set contains about $1200$  5-input canonical NPN representatives with 20 or more occurrence in IWLS 2005 benchmarks. Parameter $u$ is an integer indicating the acceptable number of uncovered practical NPN classes; $n_\text{max}$ is an integer indicating the limit number of nodes when a node reduction is needed. In our implementation, $u$ is set to $60$, and $n_\text{max}$ is set to $10000000$.

The pseudo-code for procedure \texttt{TryNode} is shown in Algorithm~\ref{alg:node}. \texttt{TryNode} creates a node, and determines whether to put it into the node list and the circuit table. Parameter $T \in \{\text{AND}, \text{XOR}\}$ indicates whether the new gate should be an AND gate or an XOR gate. Parameter $n_0$ and $n_1$ are two fanins of the new gate.

Procedure \texttt{ReduceNodes} reduces the node list by removing the nodes that are not used in any circuit in the circuit table.

Procedure \texttt{Canonicalize} calculates the canonical form of the truth table of a given function.

In the algorithms, variables $N$, $C$ and $M$ are globally accessible. $N$ denotes the list of all nodes. $C$ is a hash map of the candidate circuits; each of its entry is a set of nodes storing the root node of candidate circuits for the NPN class of this entry. $M$ is a temporary hash map to store the currently minimum costs of all functions.

\begin{algorithm}[t]
	\caption{
		\texttt{GenerateBestCircuits}($P$, $u$, $n_\text{max}$):
		Generate candidate best circuits for a subset of NPN classes of 5-input Boolean functions.
	}
	\label{alg:gen}
\begin{algorithmic}[1]
	\STATE Add constant zero node to $N$ and $C$
	\STATE Add variable nodes to $N$
	\STATE Add node of variable 0 to $C$
	\FOR{each $i$ from 2 \TO $|N|$}
		\FOR{each $j$ from 1 \TO $i - 1$}
			\STATE \texttt{TryNode}(AND, $N_i$, $N_j$)
			\STATE \texttt{TryNode}(AND, \texttt{Not}($N_i$), $N_j$)
			\STATE \texttt{TryNode}(AND, $N_i$, \texttt{Not}($N_j$))
			\STATE \texttt{TryNode}(AND, \texttt{Not}($N_i$), \texttt{Not}($N_j$))
			\STATE \texttt{TryNode}(XOR, $N_i$, $N_j$)
			\IF{num. of uncovered practical NPN classes $\leq u$}
				\RETURN
			\ENDIF
			\IF{$|N| > n_\text{max}$}
				\STATE \texttt{ReduceNodes}()
				\STATE $i \gets 1$
				\STATE break
			\ENDIF
		\ENDFOR
	\ENDFOR
\end{algorithmic}
\end{algorithm}

\begin{algorithm}[h]
	\caption{
		\texttt{TryNode}($T$, $n_0$, $n_1$):
		Create a node of type $T$ with fanins $n_0$ and $n_1$, and determine whether to put it into $N$ or $C$.
	}
	\label{alg:node}
\begin{algorithmic}[1]
	\STATE $n_\text{new} \gets$ \texttt{CreateNode}($T$, $n_0$, $n_1$)
	\STATE $t \gets \texttt{GetTruth}(n_\text{new})$
	\IF{$M_t$ not exist \OR $M_t > \texttt{Cost}(n_\text{new})$}
		\STATE $M_t \gets \texttt{Cost}(n_\text{new})$
	\ELSE
		\RETURN
	\ENDIF
	\STATE $t_\text{canon} \gets \texttt{Canonicalize}(t)$
	\IF{$\exists n \in C_{t_\text{canon}}$ such that $\texttt{Cost}(n) < \texttt{Cost}(n_\text{new})$}
		\RETURN
	\ENDIF
	\STATE add $n_\text{new}$ to the end of list $N$
	\IF{$t \neq t_\text{canon}$ \AND $t \neq \texttt{Complement}(t_\text{canon})$}
		\RETURN
	\ENDIF
	\IF{$\exists n \in C_{t_\text{canon}}$ such that $\texttt{Cost}(n) > \texttt{Cost}(n_\text{new})$}
		\STATE $C_{t_\text{canon}} \gets \emptyset$
	\ENDIF
	\IF{$t = t_\text{canon}$}
		\STATE $C_{t_\text{canon}} \gets C_{t_\text{canon}} \bigcup \{n_\text{new}\}$
	\ELSE
		\STATE $C_{t_\text{canon}} \gets C_{t_\text{canon}} \bigcup \{\texttt{Not}(n_\text{new})\}$
	\ENDIF
	\RETURN
\end{algorithmic}
\end{algorithm}

\subsection{Cut enumeration and replacement}
We use a quite similar cut enumeration and replacement technique as in~\cite{rwr}. The main difference is that we use a Boolean matcher to calculate the canonical form of the NPN representative as well as the transformation to the canonical form from the original function, while in~\cite{rwr}, a faster table look-up is used. 

The Boolean matcher proposed in~\cite{Chai06} calculates only the canonical form representation. We modified the program so it can simultaneously generate the NPN transformation, which is needed when connecting the replacement graph to the whole circuit.

Nodes are traversed in topological order. For each node starting from the PIs to the POs, all of its 5-input cuts are listed~\cite{Cong99}. The canonical form truth table and the corresponding NPN transformation of each cut are calculated using the Boolean matcher~\cite{Chai06}. Each cut is then evaluated whether there is a suitable replacement that does not increase the area of the network. Finally, the cut with the greatest gain is replaced by a best circuit. In the presented algorithm, zero-cost replacement is accepted, since it is a useful approach for re-arranging AIG structure to create more opportunities in subsequent rewriting~\cite{Bjesse04}.

The pseudo-code of the rewriting procedure is shown in Algorithm~\ref{alg:rwr}. For each node in the network, $N_\text{best}$ denotes the largest number of nodes saved by replacing a cut of the node by a pre-computed candidate circuit; $c_\text{best}$ and $u_\text{best}$ denotes the corresponding candidate circuit and the original cut, respectively. These three variables are updated simultaneously, if there exists a possible replacement.

Procedure $\texttt{ConnectToLeaves}(N, c, u, {Trans})$ connects the fanins of candidate circuit $c$ to the leaves of cut $u$, following the NPN transformation ${Trans}$.

Procedure $\texttt{Reference}(N, c)$ increases the reference count of the nodes belong to sub-circuit $c$, in network $N$, whereas $\texttt{Dereference}(N, c)$ decreases the reference count. When the reference count of a node becomes zero, the node does not belong to the network.

\begin{algorithm}[h]
	\caption{
		\texttt{RewriteNetwork}($N$, $C$):
		Rewrite a Boolean network $N$ using candidate circuits stored in hash map $C$.
	}
	\label{alg:rwr}
\begin{algorithmic}[1]
	\FOR{each node $n$ in $N$, in topological order}
		\STATE $N_\text{best} \gets -1$
		\STATE $c_\text{best} \gets \text{NULL}$
		\STATE $u_\text{best} \gets \text{NULL}$
		\FOR{each 5-input cut $u$ of $n$}
			\STATE $t \gets \texttt{GetTruth}(u)$
			\STATE $(t_\text{canon},{Trans}) \gets \texttt{Canonicalize}(t)$
			\FOR{each candidate circuit $c$ in $C_{t_\text{canon}}$}
				\STATE $\texttt{ConnectToLeaves}(N, c, u, {Trans})$
				\STATE $N_\text{saved} \gets \texttt{Dereference}(N, u)$
				\STATE $N_\text{added} \gets \texttt{Reference}(N, c)$
				\STATE $N_\text{gain} \gets N_\text{saved} - N_\text{added}$
				\STATE $\texttt{Dereference}(N, c)$
				\STATE $\texttt{Reference}(N, u)$
				\IF{$N_\text{gain} \geq 0$ and $N_best < N_\text{gain}$}
					\STATE $N_\text{best} \gets N_\text{gain}$
					\STATE $c_\text{best} \gets c$
					\STATE $u_\text{best} \gets u$
				\ENDIF
			\ENDFOR
		\ENDFOR
		\IF{$N_\text{best} = -1$}
			\STATE continue
		\ENDIF
		\STATE $\texttt{Dereference}(N, u_\text{best})$
		\STATE $\texttt{Reference}(N, c_\text{best})$
	\ENDFOR
\end{algorithmic}
\end{algorithm}

In~\cite{rwr}, the authors proposed an optimization flow composed of \emph{balance}, \emph{rewrite} and \emph{refactor} processes, and implemented it in the tool ABC~\cite{abc} with the script \emph{resyn2}. Compared to~\cite{rwr}, rewriting using 5-input cuts exploits larger cuts and more replacement options, thus has the potential for getting \emph{resyn2} script out of local minima, providing better rewriting opportunities. 

%------------------------------------------------------------------------- 
\section{Experimental Results} \label{sec:exp}

The presented algorithm is implemented using structurally hashed AIG as an internal circuit representation and integrated in ABC synthesis tool
as a command \emph{rewrite5}. 
%This command performs rewriting with 5-input cuts on the current network.
To evaluate its effectiveness, we performed a set of experiments using IWLS 2005 benchmarks~\cite{iwls2005} with more than 5000 AIG nodes after structural hashing. All experiments were carried out on a laptop with Intel Core i7 1.6GHz (2.8GHz maximum frequency) quad-core processor, 6 MB cache, and 4 GB RAM. 

First, for each benchmark, we applied a sequence of commands  \emph{resyn2; rewrite5; resyn2} in the modified ABC and compared the result to two consecutive runs of \emph{resyn2} without \emph{rewrite5} in between. 

The results are summarized in Table~\ref{tab:exp1}. Columns labeled by $A$ give the area in terms of AIG nodes. Columns labeled by $t$ give the runtime. The improvement of area and the increase of runtime are then calculated and shown in the last two columns.

Table~\ref{tab:exp1} shows that the average improvement in area achieved by adding \emph{rewrite5} in between two \emph{resyn2} runs is 3.50\%, at the cost of 33.18\% of extra runtime. This result indicates that the proposed \emph{rewrite5} method is effective in bringing ABC's \emph{resyn2} optimization script out of local minima, leading to better optimization possibilities.

The second experiment is performed similarly, except we used a longer optimization flow: \emph{resyn2; rewrite5; resyn2; rewrite5; resyn2}. The result is compared to three consecutive runs of \emph{resyn2} script.

The result of the second experiment is shown in Table~\ref{tab:exp2}, which has the same structure as Table~\ref{tab:exp1}. The average improvement in area using the new optimization flow is 4.88\%, at the cost of 46.11\% of extra runtime. This result shows the possibility to further extend the \emph{resyn2} sequence by inserting \emph{rewrite5} runs, to achieve even better optimization.

Even longer optimization flows were also tested. The comparison of average results is summarized in Table~\ref{tab:sum}. The improvement in area converges after certain number of \emph{resyn2}-\emph{rewrite5} iterations. The increase of improvement is insignificant for more than four runs of \emph{resyn2}. 

\begin{table*}[p]
	\centering
	\footnotesize
\begin{tabular}{cr|rr|rr|rr}
\hline
           &            & \multicolumn{ 2}{|c}{resyn2;resyn2} & \multicolumn{ 2}{|c|}{resyn2;rewrite5;resyn2} &            &            \\
\cline{3-6}
 benchmark &      nodes &     $A_1$ & $t_1$, sec &     $A_2$ & $t_2$, sec & $(A_1-A_2)/A_1$ & $(t_2-t_1)/t_1$ \\
\hline
ac97\_ctrl &      14244 &      10222 &     0.759  &      10212 &     0.921  &     0.10\% &    21.34\% \\

 aes\_core &      21522 &      20153 &     3.125  &      19945 &     4.079  &     1.03\% &    30.53\% \\

    b14\_1 &       9471 &       5902 &     1.299  &       4712 &     1.929  &    20.16\% &    48.50\% \\

    b15\_1 &      17015 &      10215 &     2.067  &      10012 &     2.204  &     1.99\% &     6.63\% \\

    b17\_1 &      51419 &      31447 &     5.364  &      30943 &     6.948  &     1.60\% &    29.53\% \\

    b18\_1 &     130418 &      81185 &    18.947  &      78430 &    25.344  &     3.39\% &    33.76\% \\

    b19\_1 &     254960 &     153796 &    37.618  &     149269 &    47.708  &     2.94\% &    26.82\% \\

    b20\_1 &      21074 &      13635 &     2.666  &      12048 &     3.819  &    11.64\% &    43.25\% \\

    b21\_1 &      20538 &      12845 &     2.618  &      10940 &     3.900  &    14.83\% &    48.97\% \\

    b22\_1 &      31251 &      19698 &     4.109  &      16986 &     5.870  &    13.77\% &    42.86\% \\

 des\_perf &      82650 &      73724 &    15.717  &      73224 &    23.228  &     0.68\% &    47.79\% \\

       DMA &      24389 &      22306 &     2.524  &      20269 &     3.129  &     9.13\% &    23.97\% \\

       DSP &      44759 &      37976 &     5.635  &      37728 &     7.734  &     0.65\% &    37.25\% \\

  ethernet &      86650 &      55925 &     5.790  &      55838 &     7.879  &     0.16\% &    36.08\% \\

     leon2 &     788737 &     774919 &   142.645  &     774065 &   187.660  &     0.11\% &    31.56\% \\

 mem\_ctrl &      15325 &       8518 &     1.255  &       8449 &     1.511  &     0.81\% &    20.40\% \\

   netcard &     803723 &     516124 &    93.952  &     516001 &   122.749  &     0.02\% &    30.65\% \\

pci\_bridge32 &      22790 &      16362 &     1.719  &      16271 &     2.288  &     0.56\% &    33.10\% \\

    s35932 &       8371 &       7843 &     0.755  &       7843 &     1.003  &     0.00\% &    32.85\% \\

    s38417 &       9062 &       7969 &     0.812  &       7936 &     1.149  &     0.41\% &    41.50\% \\

    s38584 &       8477 &       7224 &     0.720  &       7188 &     0.921  &     0.50\% &    27.92\% \\

systemcaes &      12384 &       9614 &     1.705  &       9391 &     2.602  &     2.32\% &    52.61\% \\

      tv80 &       9635 &       7084 &     1.169  &       6970 &     1.498  &     1.61\% &    28.14\% \\

usb\_funct &      15826 &      13082 &     1.439  &      12892 &     1.858  &     1.45\% &    29.12\% \\

  vga\_lcd &     126696 &      88641 &    10.517  &      88659 &    14.268  &    -0.02\% &    35.67\% \\

wb\_conmax &      47853 &      39163 &     4.748  &      38701 &     5.791  &     1.18\% &    21.97\% \\
\hline
   Average &            &            &            &            &            &     3.50\% &    33.18\% \\
\hline
\end{tabular}  

	\caption{Effectiveness of improving double \emph{resyn2} optimization flow using \emph{rewrite5}, on IWLS 2005 benchmarks.}
	\label{tab:exp1}
\end{table*}

\begin{table*}[p]
	\centering
	\footnotesize
	
\begin{tabular}{cr|rr|rr|rr}
\hline
           &            & \multicolumn{ 2}{|c}{resyn2;resyn2;resyn2} & \multicolumn{ 2}{|p{25mm}|}{resyn2;rewrite5;resyn2; rewrite5;resyn2} &            &            \\
\cline{3-6}
 benchmark &      nodes &     $A_1$ & $t_1$, sec &     $A_2$ & $t_2$, sec & $(A_1-A_2)/A_1$ & $(t_2-t_1)/t_1$ \\
\hline
ac97\_ctrl &      14244 &      10202 &     1.084  &      10180 &     1.396  &     0.22\% &    28.78\% \\

 aes\_core &      21522 &      20044 &     4.562  &      19554 &     6.646  &     2.44\% &    45.68\% \\

    b14\_1 &       9471 &       5652 &     1.702  &       4350 &     2.526  &    23.04\% &    48.41\% \\

    b15\_1 &      17015 &      10029 &     2.335  &       9796 &     3.231  &     2.32\% &    38.37\% \\

    b17\_1 &      51419 &      30107 &     7.446  &      29248 &    10.530  &     2.85\% &    41.42\% \\

    b18\_1 &     130418 &      79204 &    24.658  &      74827 &    38.047  &     5.53\% &    54.30\% \\

    b19\_1 &     254960 &     149177 &    49.815  &     143633 &    70.876  &     3.72\% &    42.28\% \\

    b20\_1 &      21074 &      13405 &     3.811  &      10732 &     5.878  &    19.94\% &    54.24\% \\

    b21\_1 &      20538 &      12240 &     3.603  &       9379 &     5.437  &    23.37\% &    50.90\% \\

    b22\_1 &      31251 &      18967 &     5.614  &      15186 &     8.595  &    19.93\% &    53.10\% \\

 des\_perf &      82650 &      73248 &    23.235  &      72322 &    36.941  &     1.26\% &    58.99\% \\

       DMA &      24389 &      22288 &     3.573  &      20214 &     4.874  &     9.31\% &    36.41\% \\

       DSP &      44759 &      37634 &     8.055  &      37273 &    12.465  &     0.96\% &    54.75\% \\

  ethernet &      86650 &      55803 &     8.287  &      55794 &    12.067  &     0.02\% &    45.61\% \\

     leon2 &     788737 &     774560 &   213.921  &     773399 &   352.054  &     0.15\% &    64.57\% \\

 mem\_ctrl &      15325 &       8408 &     1.726  &       8313 &     2.260  &     1.13\% &    30.94\% \\

   netcard &     803723 &     515961 &   133.294  &     515771 &   181.877  &     0.04\% &    36.45\% \\

pci\_bridge32 &      22790 &      16313 &     2.385  &      16235 &     3.650  &     0.48\% &    53.04\% \\

    s35932 &       8371 &       7843 &     1.034  &       7843 &     1.457  &     0.00\% &    40.91\% \\

    s38417 &       9062 &       7947 &     1.158  &       7886 &     1.725  &     0.77\% &    48.96\% \\

    s38584 &       8477 &       7217 &     1.021  &       7199 &     1.312  &     0.25\% &    28.50\% \\

systemcaes &      12384 &       9595 &     2.258  &       9248 &     4.043  &     3.62\% &    79.05\% \\

      tv80 &       9635 &       7030 &     1.618  &       6879 &     2.308  &     2.15\% &    42.65\% \\

usb\_funct &      15826 &      13041 &     2.037  &      12784 &     2.880  &     1.97\% &    41.38\% \\

  vga\_lcd &     126696 &      88621 &    15.258  &      88687 &    22.223  &    -0.07\% &    45.65\% \\

wb\_conmax &      47853 &      38676 &     6.759  &      38095 &     9.032  &     1.50\% &    33.63\% \\
\hline
   Average &            &            &            &            &            &     4.88\% &    46.11\% \\
\hline
\end{tabular}  

	\caption{Effectiveness of improving triple \emph{resyn2} optimization flow using \emph{rewrite5}, on IWLS 2005 benchmarks.}
	\label{tab:exp2}
\end{table*}

\begin{table}[h]
	\centering
	\footnotesize
\begin{tabular}{c|cc}
\hline
           & improvement in area & extra runtime \\
\hline
 SS $\rightarrow$ SWS &     3.50\% &    33.18\% \\

SSS $\rightarrow$ SWSWS &     4.88\% &    46.11\% \\

SSSS $\rightarrow$ SWSWSWS &     5.39\% &    47.48\% \\

SSSSS $\rightarrow$ SWSWSWSWS &     5.57\% &    51.21\% \\
\hline
\multicolumn{2}{l}{NOTE: S stands for \emph{resyn2}; W stands for \emph{rewrite5}.}
\end{tabular}  
	\caption{Summary of average results.}
	\label{tab:sum}
\end{table}

%------------------------------------------------------------------------- 
\section{Conclusion} \label{sec:conc}

In this paper, we present an AIG-based rewriting technique that uses 5-input cuts. The technique extends the approach of AIG rewriting using 4-input cuts presented in~\cite{rwr}. Experimental results show that our algorithm is effective in driving other optimization techniques, such as \emph{resyn2} script in ABC, out of local minima. The proposed rewriting technique might be useful in a new optimization flow combining rewriting of both 4-input and 5-input cuts.

%------------------------------------------------------------------------- 
\balance

\bibliographystyle{IEEEtran}
\bibliography{bib}

\end{document}